\documentclass[aps,physrev,preprint,superscriptaddress,showkeys,floatfix]{revtex4-2}

\usepackage{graphicx,amsmath,longtable} 
\usepackage[T1]{fontenc}
\usepackage{xr}
\begin{document}


\title{Ultrafast heat transfer in single palladium nanocrystals seen with an X-ray free-electron laser}

\author{David Yang}
\email{dyang2@bnl.gov}
\affiliation{Condensed Matter Physics and Materials Science Department, Brookhaven National Laboratory, Upton, NY 11973, USA}

\author{James Wrigley}
\affiliation{European X-Ray Free-Electron Laser Facility, Holzkoppel 4, 22869 Schenefeld, Germany}

\author{Jack Griffiths}
\affiliation{Condensed Matter Physics and Materials Science Department, Brookhaven National Laboratory, Upton, NY 11973, USA}

\author{Longlong Wu}
\altaffiliation[Present address: ]{Shanghai Advanced Research Institute, Chinese Academy of Sciences, Shanghai 201210, China.}
\affiliation{Condensed Matter Physics and Materials Science Department, Brookhaven National Laboratory, Upton, NY 11973, USA}

\author{Ana F. Suzana}
\altaffiliation[Present address: ]{Chemical Sciences and Engineering Division, Argonne National Laboratory, Lemont, IL, 60439, United States}
\affiliation{Condensed Matter Physics and Materials Science Department, Brookhaven National Laboratory, Upton, NY 11973, USA}

\author{Jiecheng Diao}
\altaffiliation[Present address: ]{Center for Transformative Science, ShanghaiTech University, Shanghai 201210, China.}
\affiliation{London Centre for Nanotechnology, University College London, London WC1E 6BT, United Kingdom.}

\author{Angel Rodriguez-Fernandez}
\affiliation{European X-Ray Free-Electron Laser Facility, Holzkoppel 4, 22869 Schenefeld, Germany}

\author{Joerg Hallmann}
\affiliation{European X-Ray Free-Electron Laser Facility, Holzkoppel 4, 22869 Schenefeld, Germany}

\author{Alexey Zozulya}
\affiliation{European X-Ray Free-Electron Laser Facility, Holzkoppel 4, 22869 Schenefeld, Germany}

\author{Ulrike Boesenberg}
\affiliation{European X-Ray Free-Electron Laser Facility, Holzkoppel 4, 22869 Schenefeld, Germany}

\author{Roman Shayduk}
\affiliation{European X-Ray Free-Electron Laser Facility, Holzkoppel 4, 22869 Schenefeld, Germany}


\author{Jan-Etienne Pudell}
\affiliation{European X-Ray Free-Electron Laser Facility, Holzkoppel 4, 22869 Schenefeld, Germany}

\author{Anders Madsen}
\affiliation{European X-Ray Free-Electron Laser Facility, Holzkoppel 4, 22869 Schenefeld, Germany}



\author{Ian K. Robinson}
\email{i.robinson@ucl.ac.uk}
\affiliation{Condensed Matter Physics and Materials Science Department, Brookhaven National Laboratory, Upton, NY 11973, USA}
\affiliation{London Centre for Nanotechnology, University College London, London WC1E 6BT, United Kingdom.}

\date{\today}

\begin{abstract}
We report transient highly strained structural states in individual palladium (Pd) nanocrystals, electronically heated using an optical laser, which precede their uniform thermal expansion. Using an X-ray free-electron laser probe, the evolution of individual $111$ Bragg peaks is measured as a function of delay time at various laser fluences. Above a laser fluence threshold at a sufficient pump-probe delay, the Bragg peak splits into multiple peaks, indicating heterogeneous strain, before returning to a single peak, corresponding to even heat distribution throughout the lattice expanded crystal. Our findings are supported by a lattice displacement and strain model of a single nanocrystal at different delay times, which agrees with the experimental data. Our observations have implications for understanding femtosecond laser interactions with metals and the potential photo-catalytic performance of Pd.
\end{abstract}


\keywords{palladium, coherent x-ray diffraction imaging, nanocrystal, heat transfer}

\maketitle
\newpage

\section{\label{sec:Introduction}Introduction}
The advent of X-ray free-electron lasers (XFELs) has revolutionised the study of ultrafast structural dynamics in materials: hard X-rays are intense enough to measure diffraction from single nanocrystals in a single pulse \cite{Xu2014}. Metallic nanocrystals are intriguing because they exhibit unique electronic, optical, and catalytic properties governed by their size, shape, surfaces, and lattice dynamics \cite{Link2003}. Palladium (Pd) nanocrystals, in particular, play a crucial role in catalysis, hydrogen storage, medicine, and plasmonics \cite{Joudeh2022}. Yet, their non-equilibrium structural response to optical excitation remains poorly understood at ultrafast timescales and has only been addressed using molecular dynamics (MD) simulations \cite{Kateb2018}. For Pd, this non-equilibrium structural response affects energy dissipation pathways, which are crucial for optimising photo-catalytic applications \cite{Li2020d}.


Optical excitation of metals triggers rapid electron thermalisation on a femtosecond (fs) timescale, followed by electron-phonon coupling that transfers energy to the lattice over picoseconds (ps) \cite{Hohlfeld2000}. By combining femtosecond X-ray pulses with optical laser excitation, stroboscopic ``pump-probe'' techniques can capture transient structural changes in metallic nanocrystals on picosecond timescales. This methodology has been critical for understanding phenomena such as coherent phonon oscillations \cite{Clark2013}, and for sufficiently high heating rates, lattice melting \cite{Assefa2020a,Antonowicz2024,Suzana2023b,Clark2015a,Ihm2019,Sun2025}. Moreover, ultrafast X-ray diffraction enables the study of thin film heterostructures, revealing heat transfer pathways between metals with different electron-phonon coupling parameters \cite{Pudell2020, Pudell2018}. 

Several other ultrafast time-resolved methods have been used to study heat transfer in metals. Time-resolved X-ray absorption near-edge structure (XANES) on a copper thin film has revealed the loss of crystallinity on the order of 1 ps pump-probe delay \cite{Jourdain2021}. Optical reflectivity measurements on aluminium (Al) thin films demonstrate the transfer of heat from the photoexcited electrons to the lattice on the order of 2 ps delay \cite{Kandyla2007}. Al thin films have also been studied using femtosecond electron diffraction, revealing a short-range order liquid structure after 3.5 ps delay under a laser fluence of $\mathrm{70\ mJ/cm^2}$, providing evidence for electron-phonon coupling-mediated melting \cite{Siwick2003}. Ultrafast electron diffraction has also been applied to gold (Au) thin films to reveal heterogeneous to homogeneous melting pathways \cite{Mo2018}. 

\begin{figure}[ht]
    \centering
    \includegraphics[width=0.6\linewidth]{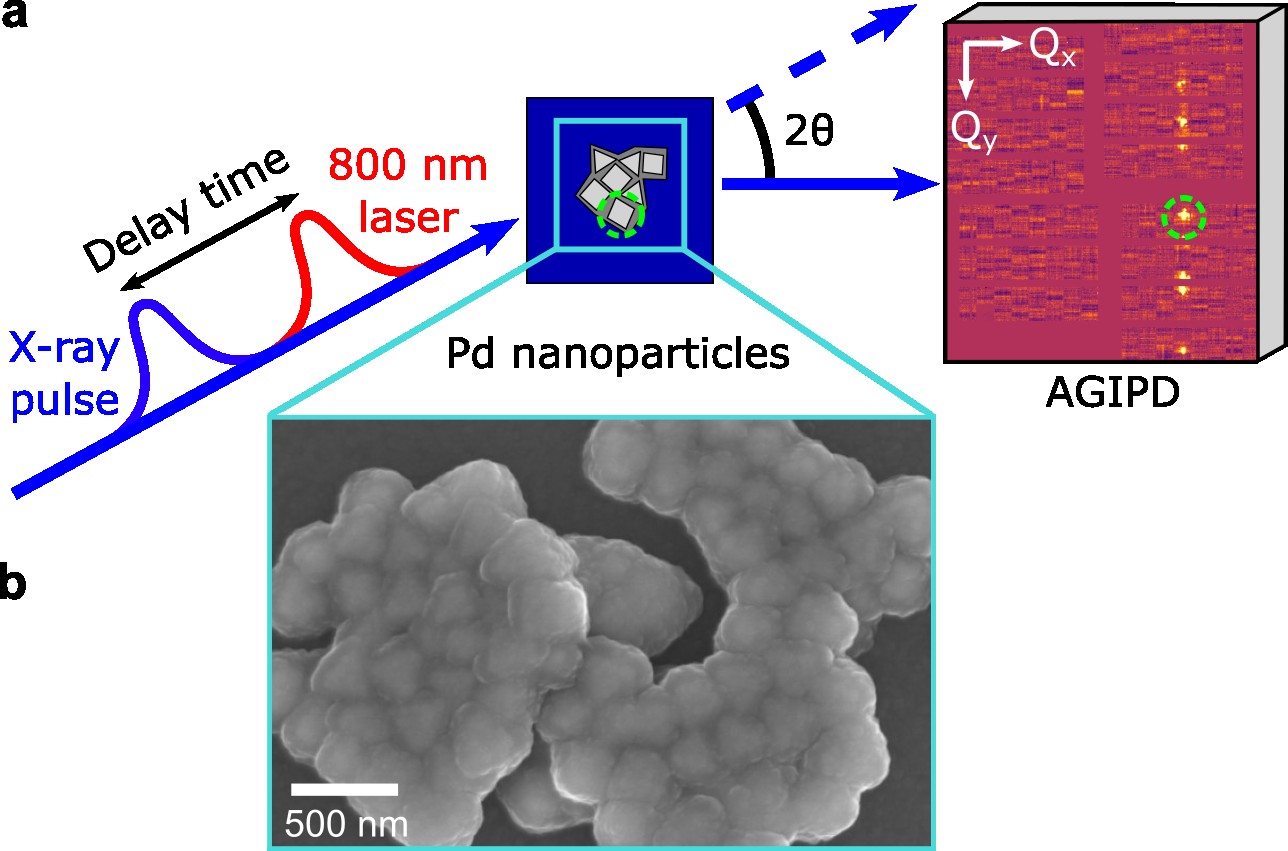}
    \caption{\textbf{Experimental ultrafast pump-probe X-ray diffraction setup in horizontal scattering geometry.} \textbf{a} The samples are pumped using an 800 nm optical laser, followed by an X-ray probe with a variable delay time. An isolated $111$ Bragg peak, corresponding to a single Pd nanoparticle (green dashed circle), was measured using the adaptive gain integrating pixel detector (AGIPD) positioned 4 m from the sample at $2\theta$ = 35.8 °. White arrows indicate the $Q_x$ and $Q_y$ directions in reciprocal space. \textbf{b} SEM image of octahedral-shaped Pd nanocrystals coated with $\approx 10$ nm of $\mathrm{TiO_2}$ for stability. The lighter, triangle-shaped profiles show the Pd particle facets, surrounded by the darker $\mathrm{TiO_2}$ coating. These coated nanoparticles are later fixed on a Si substrate using tetraethyl orthosilicate (TEOS), which is not shown.}
    \label{fig:SEM}
\end{figure}

Here, we probe the transient structural states of Pd nanocrystals excited by an 800 nm optical laser (Fig. \ref{fig:SEM}). Two-dimensional (2D) $111$ coherent X-ray diffraction patterns from individual nanoparticles were measured at various delay times and laser fluences ranging from $\mathrm{57 - 230\ mJ/cm^2}$. By monitoring the Bragg peak positions, we can infer that a macroscopic, out-of-plane rotation of the crystal takes place after pumping. Using coherent imaging results, we generate a distinct model of the heterogeneous lattice strain distribution within one of the crystals, which we attribute to the inhomogeneous distribution of electrons generated by the laser. This model illustrates the build-up of a strain distribution with simultaneously compressed and expanded regions within a single nanocrystal, lasting approximately 20 ps.

The two-temperature model (TTM) has been frequently used to describe the response of a material to a femtosecond laser pulse \cite{Anisimov1974}. The laser heats electrons on the surface of a metal crystal to a very high temperature. This heat is then transferred to the crystal lattice first by electron diffusion, and then by electron-phonon coupling to the lattice over a few picoseconds. The spatial distribution of the heat transfer throughout the crystal depends on the mean free path of electrons, which is both temperature and material-dependent \cite{Chen2006}. The leading model for the melting of bulk polycrystalline materials is a heterogeneous to homogeneous pathway \cite{Lin2006,Mo2018,Antonowicz2024}, involving atoms at defects, surfaces, and grain boundaries becoming disordered first, followed by the rest of the crystal volume at the timescale dictated by thermal diffusion. On which timescales these events take place in a single nanocrystal is addressed by the results of the experiments described here: the electrons get optically heated and travel only partway through the crystal, upon which they then transfer heat to the lattice and cause heterogeneous strain.

Previous studies have demonstrated that intense optical excitation induces non-thermal melting, especially in semiconductors \cite{Sokolowski-Tinten2001}. In thin polycrystalline Pd films, the transition from crystalline to liquid states proceeds heterogeneously via grain boundaries before evolving into homogeneous melting at higher fluences \cite{Lin2006,Mo2018,Suzana2023b, Antonowicz2024,Olczak2025}. Lattice compression lasting tens of picoseconds, found in Pd \cite{Suzana2023b} and Pt \cite{Gelisio2024} films, is an important signature of inhomogeneity of the thermal distribution \cite{Thomsen1986}.  A similar study performed on Au thin films \cite{Assefa2020a} did not reveal any compressive response, while this was seen in Cu \cite{Milathianaki2013} at high laser fluences of a long-pulse drive laser. Unlike the previous ensemble-averaged measurements, individual Au nanocrystals have been studied using similar optical laser pump-XFEL probe experiments, and did not show any signs of compression \cite{Clark2013, Clark2015a}. By analogy with the thin film results, the melting pathway of individual Pd nanocrystals reported in this work is expected to differ from Au nanocrystals since the electron-phonon coupling rate is greater for Pd \cite{Medvedev2020}.




\section{\label{sec:Results}Results}

\subsection{\label{sec:Bragg peak evolution}Bragg peak evolution}
We first report the pump-probe X-ray diffraction patterns for three Pd nanoparticles, labelled as crystals A-C, at different optical laser fluences and delay times. Different crystals were selected from the composite shown in Fig.\ref{fig:SEM}b. Under the experimental conditions (Methods), we assume the measured crystals were near the surface of the composite that faces the laser, where heating predominantly occurs. 

\begin{figure*}[ht]
    \centering
    \includegraphics[width=\linewidth]{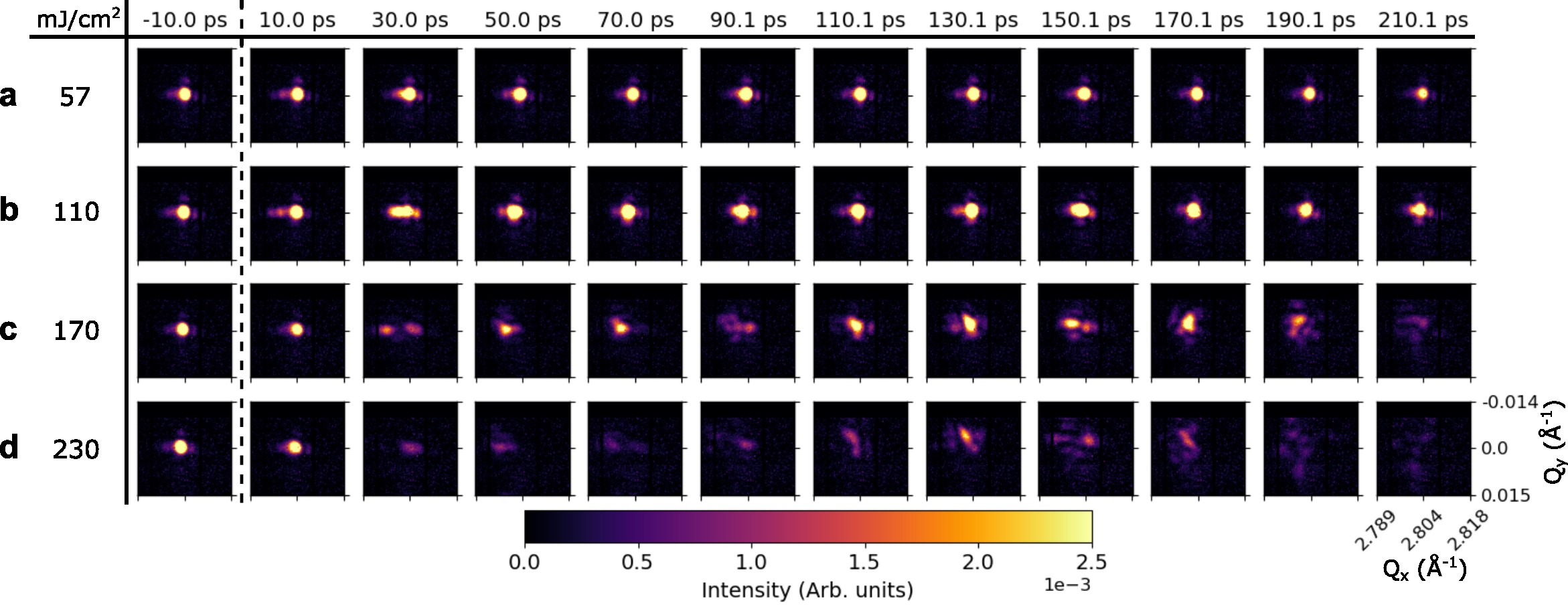}
    \caption{\textbf{Evolution of the $111$ Bragg peak intensity on a linear scale for crystal A at various delay times and laser fluences.} Rows \textbf{a} - \textbf{d} correspond to sequential delay measurements using different fluences. The evolution of crystals B and C is shown in Supplementary Figs. \ref{supp-fig:Bragg_peaks_B} - \ref{supp-fig:Bragg_peaks_C-2}.}
    \label{fig:Bragg_peaks}
\end{figure*}

The evolution of the $111$ 2D Bragg peak for crystal A at selected delay times is shown in Fig. \ref{fig:Bragg_peaks}. The optical laser pump was timed to coincide with alternating collinear X-ray probe pulses of the XFEL. Analogous data for crystals B and C are in Supplementary Figs. \ref{supp-fig:Bragg_peaks_B} - \ref{supp-fig:Bragg_peaks_C-2}. Before the measurement, the diffraction was aligned to the maximum of its rocking curve without laser pumping. For crystal A, the measurements were performed sequentially, from low to high laser fluence, alternating between laser-pumped and unpumped XFEL pulses (Methods). Unpumped pulses for crystal A, shown in Supplementary Fig. \ref{supp-fig:Bragg_peaks_A_unpumped}, proved that the crystal returns to its original state after each laser pulse at all fluences.

At the lowest measured laser fluence, $\mathrm{57\ mJ/cm^2}$, there were slight horizontal translations of the Bragg peak position on the detector, denoting changes in the Bragg angle ($\theta$) but the peak shape was relatively unchanged. At a laser fluence of $\mathrm{110\ mJ/cm^2}$, we observed a slight splitting of the Bragg peak at 30 ps delay time, which then resumes as a single peak after 50 ps, though shifted in peak position ($Q_{x,y}$). This was followed by relative peak translations and some Bragg peak disorder, relative to the lowest laser fluence. For the next laser fluence, $\mathrm{170\ mJ/cm^2}$, there was a more pronounced peak splitting and disorder, following the same timeline as observed for $\mathrm{110\ mJ/cm^2}$. This clear splitting is not apparent at later delay times, which we attribute to damping \cite{Cavalleri2000}. However, we note a strong decrease in Bragg peak intensity, particularly at 90 ps and 210 ps. A drop in Bragg peak intensity at positive delay times is most evident at $\mathrm{230\ mJ/cm^2}$.

Small vertical peak shifts can be interpreted as the rotation of the crystals \cite{Diao2020a}, while small peak shifts along the horizontal direction ($2\theta$) are due to changes in the average lattice parameter. The change in average lattice relative to its unperturbed state at negative delay times represents a homogeneous strain. Heterogeneous strain, relative to the average lattice, can be inferred from the changes in the full width at half maximum (FWHM) of the Bragg peak line profile. The FWHM provides information about the local distortion of the crystal lattice.

\subsection{\label{sec:Incoherent diffraction analysis}Incoherent imaging}

\begin{figure}[ht]
    \centering
    \includegraphics[width=\linewidth]{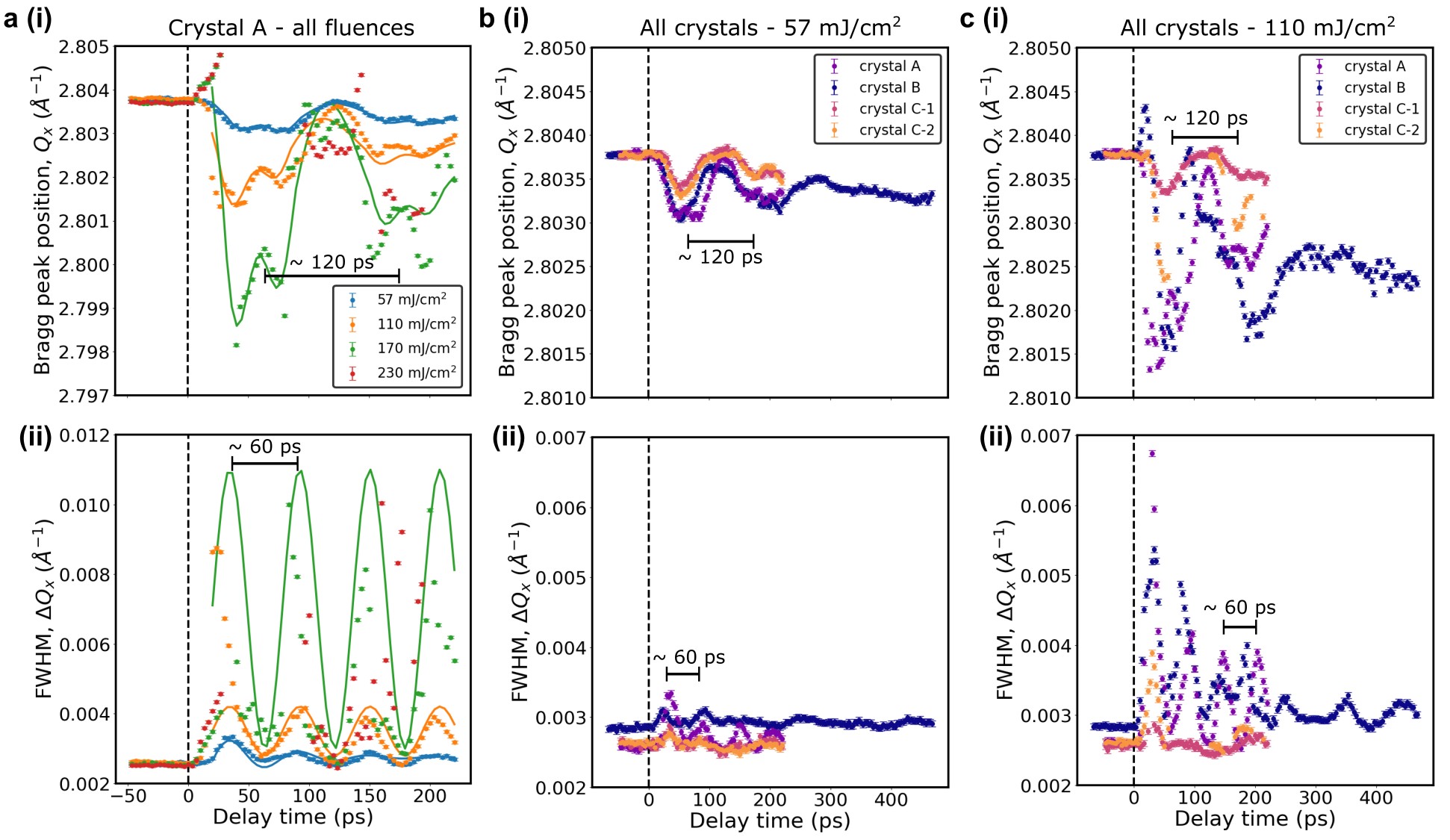}
    \caption{\textbf{Summary of the changes to the Gaussian-fitted Bragg peaks along the horizontal $2\theta$ direction, $Q_x$, as a function of laser fluence and delay time.} \textbf{a} Fitted parameters for various laser fluences for Crystal A. (i) Fitted position of the Bragg peak along $Q_x$. The oscillations were fitted with Eq. \ref{supp-eq:peak_position_fitting} for all fluences except for $\mathrm{230\ mJ/cm^2}$ due to a lack of signal. (ii) Fitted FWHM of the Bragg peak along $Q_x$. The oscillations were fitted with Eq. \ref{supp-eq:peak_FWHM_fitting}. The error of the Gaussian fits is shown in Supplementary Fig. \ref{supp-fig:Crystal_A_chi2_error}. \textbf{b} Fitted parameters for a laser fluence of $\mathrm{57\ mJ/cm^2}$ on all measured crystals. \textbf{c} Fitted parameters for a laser fluence of $\mathrm{110\ mJ/cm^2}$ on all measured crystals. Crystal C-1 and C-2 are two successive measurements on the same crystal. The error bars reflect two standard deviations of repeated measurements at negative delay times. Oscillation periods are shown by the horizontal bars in each panel. The error of the Gaussian fits is shown in Supplementary Fig. \ref{supp-fig:Fluence_chi2_error} for \textbf{b} and \textbf{c}. The equivalent of this figure in the $Q_y$ direction is Supplementary Fig. \ref{supp-fig:Bragg_peak_COM_FWHM_y}.} 
    \label{fig:Bragg_peak_COM_FWHM_x}
\end{figure}

To quantify the evolution of the Bragg peaks, they were approximated as 2D Gaussian functions (Methods). The fitted peak positions and FWHM along the horizontal detector direction, $Q_x$, are plotted as a function of delay time in Fig. \ref{fig:Bragg_peak_COM_FWHM_x}(i). Even though the Bragg peak exhibits clear splitting above $\mathrm{110\ mJ/cm^2}$ at 30 ps delay time (Fig. \ref{fig:Bragg_peaks}), we approximated the peak positions and FWHM using single-peaked Gaussian functions. Note the $\chi^2$ error of the Gaussian fitting for crystal A is shown in Supplementary Fig. \ref{supp-fig:Crystal_A_chi2_error} and only fits with a $\chi^2<0.3$ were considered for analysis. 

We observe an initial homogeneous compression of the crystal between 0 to 20 ps above $\mathrm{57\ mJ/cm^2}$ in Fig. \ref{fig:Bragg_peak_COM_FWHM_x}a(i) and c(i). Then, the crystal expands in an oscillatory pattern for the remainder of the delay scan. We see a prominent oscillation along the horizontal direction, fitted using Eq. \ref{supp-eq:peak_position_fitting} (see Supplementary Note \ref{supp-sec:Oscillation fitting}), showing a period of $\approx 120$ ps (Fig. \ref{fig:Bragg_peak_COM_FWHM_x}a(i)). This corresponds to crystal A's lowest-frequency vibrational mode. As the laser fluence increases, we observe an increase in the oscillation amplitude of the Bragg peak position. The increased laser fluence also causes greater homogeneous lattice expansion, as seen in the greater displacement in the horizontal direction. 

In Fig. \ref{fig:Bragg_peak_COM_FWHM_x}a-c(ii), we observe oscillations in the horizontal FWHM, $\Delta Q_x$, of the Gaussian-fitted crystals. Using the fitting function shown in Eq. \ref{supp-eq:peak_FWHM_fitting} (see Supplementary Note \ref{supp-sec:Oscillation fitting}), we observe an oscillation with a period of 60 ps, most apparent for a laser fluence of $\mathrm{110\ mJ/cm^2}$. Interestingly, this is half the period of the 120 ps vibrational mode, corresponding to the acoustic wave propagating forwards and backwards through the crystal \cite{Nicoul2011}. The horizontal FWHM is a measure of the presence of heterogeneous strain, which would be greatest when the acoustic wave lies halfway through the crystal, occurring twice per acoustic wave period. 

To test the reproducibility of our observations, we measured two other crystals, B and C, at laser fluences of $\mathrm{57\ mJ/cm^2}$ and $\mathrm{110\ mJ/cm^2}$. The Bragg peaks again were fitted as 2D Gaussian functions, and the resulting centre of masses and FWHMs are plotted as a function of delay time, shown in Fig. \ref{fig:Bragg_peak_COM_FWHM_x}b and c. While measuring only three crystals limits generality, the period of the oscillations in the horizontal direction is consistent for both laser fluences and all crystals, and the oscillation amplitudes increase for all crystals as the laser fluence increases.

The FWHM of the fitted Gaussians show similar oscillation periods along the horizontal direction, with an oscillation amplitude increasing with laser fluence. The fitted FWHM is different for each crystal, owing to slight differences in crystal morphology and initial residual strain state, but they all show similar oscillations along the horizontal direction.

The peak movements in the vertical detector direction, $Q_y$, are presented in Supplementary Fig. \ref{supp-fig:Bragg_peak_COM_FWHM_y} for all measured crystals. While no oscillations were observed, peak translations, interpreted as rotations \cite{Diao2020a}, were found to increase with laser fluence. The degree of rotation increases with laser fluence, up to $-0.003\ \textup{~\AA}^{-1}$ or $-0.5$ mrad ($\approx0.03\ ^{\circ}$) for $\mathrm{230\ mJ/cm^2}$. We attribute the notable decrease in Bragg peak intensity in Fig. \ref{fig:Bragg_peaks}d to the selected Pd crystal being stroboscopically rotated out of Bragg condition. The crystal rotation along the rocking axis as a function of delay time was estimated from the drop in intensity by calibrating against a rocking curve, shown in Supplementary Fig. \ref{supp-fig:Rocking_curve}, indicating there is a maximum rotation of $\approx0.06\ ^{\circ}$. This rotation is largely due to the lattice parameter expanding and contracting (see Supplementary Note \ref{supp-sec:Crystal rotation}). Supplementary Fig. \ref{supp-fig:Bragg_peak_COM_FWHM_y} shows that there is no consistency in the rotation direction for each crystal. This can be attributed to the local arrangement of nanocrystal neighbours \cite{Diao2020a}, which would have been different for each of our measured nanocrystals.

\subsection{\label{sec:Coherent imaging}Model of coherent diffraction data}
We now turn to the question of the details of the strain detected in the FWHM of the diffraction patterns discussed above, paying attention to the first 50 ps. The XFEL beam used for the measurements was fully coherent, thus the diffraction can be interpreted in the language of Bragg Coherent Diffraction Imaging (BCDI) \cite{Robinson2009}. The measured BCDI patterns were sufficiently oversampled by at least twice the Nyquist frequency \cite{Sayre1952}, at least 2 pixels per fringe along each dimension, so iterative phase retrieval algorithms could be used to recover the phase of the Bragg peak \cite{Fienup1982}, enabling us to obtain real-space images of the sample through the Fourier transform. The corresponding phase of the real-space image represents the projection of the displacement field onto the scattering vector, $\mathbf{u_{111}}$, from which we can also determine the corresponding strain field along the same direction, $\varepsilon_{111}$ \cite{Robinson2009}.

For our measurements, which were 2D sections of the coherent diffraction patterns, the low signal-to-noise levels rendered the attempted reconstructions non-reproducible. The diversity of solutions renders advanced phase retrieval algorithms, such as those implementing genetic algorithms \cite{Chen2007}, unsuccessful. A machine-learning-based method has been recently developed for ultrafast 2D coherent diffraction patterns \cite{Yu2024}, but was trained on phase domain structures in thin films and is inappropriate for Pd nanocrystals. 

\begin{figure*}
    \centering
    \includegraphics[width=\linewidth]{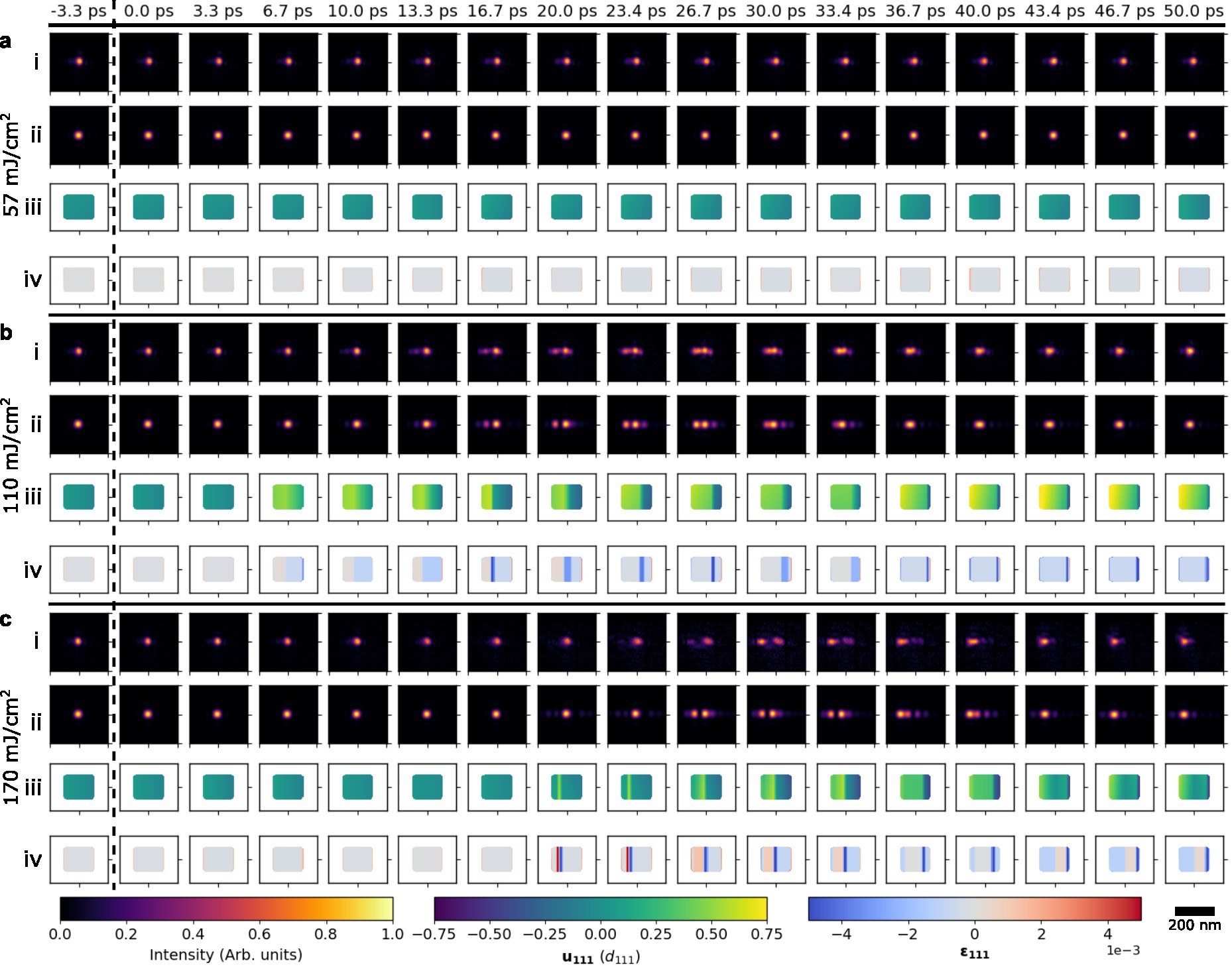}
    \caption{\textbf{Real-space 2D model of crystal A in the laboratory frame up to 50 ps delay time at \textbf{a} $\mathrm{57\ mJ/cm^2}$, \textbf{b} $\mathrm{110\ mJ/cm^2}$ and \textbf{c} $\mathrm{170\ mJ/cm^2}$.} For each subfigure, (i) the experimental Bragg peak, (ii) the model Bragg peak, (iii) the real space model displacement, $\mathbf{u_{111}}$, and (iv) the real space model strain, $\varepsilon_{111}$, are presented as rows. (i) and (ii) refer to the bottom left colour bar, (iii) refers to the bottom middle colour bar and (iv) refers to the bottom right bar. We note that $\mathbf{u_{111}}$ and $\varepsilon_{111}$ also contain the homogeneous displacement and strain that result from the Bragg peak shifts relative to negative delay times. The experimental and model Bragg peaks are both normalised for ease of comparison. The model parameters are shown in Fig. \ref{fig:Strain_disp_values}. The centre of mass parameters are shown in Supplementary Fig. \ref{supp-fig:Model_parameters}c and d. The error associated with the model is shown in Supplementary Fig. \ref{supp-fig:Model_error}.}
    \label{fig:Model}
\end{figure*}

As a simpler approach, a 1D forward model of the complex phase function describing crystal A was created and used to fit the measured Bragg peaks at various delay times and laser fluences instead (Methods). In order to explain peak splitting in the horizontal ($2\theta$ or $Q_x$) and little change in the vertical ($Q_y$) directions, we assumed a rectangular-shaped crystal (for simplicity) of uniform amplitude with a boundary located at $x=x_0$ inside it (Eq. \ref{eq:S}). Within the region $x\leq x_0$, the image phase ramps up with a positive slope, $s_1$, while in the region $x>x_0$, it ramps down with a negative slope, $s_2$ (Eq. \ref{eq:psi}). These three fit parameters were sufficient to explain the transient splitting of the peak at higher fluences in the early delay time range (< 60 ps) and also to explain the enlarged FWHM at higher fluences. The slopes correspond to the peak shifts, while $x_0$ couples to the relative peak intensity of the split peak and its asymmetry at lower fluences. Following the usual BCDI convention \cite{Robinson2009}, the image phase is proportional to the local lattice displacement along $Q$ and its derivative is the local strain in the crystal. Here, we highlight that the model $\mathbf{u_{111}}$ and $\varepsilon_{111}$ also contain the homogeneous displacement and strain that result from the Bragg peak shifts relative to negative delay times. For each fluence and delay time in Fig \ref{fig:Model}, we show images of the displacement in row (iii) and the strain in row (iv), as fitted to the BCDI data with this model. The error associated with the model is shown in Supplementary Fig. \ref{supp-fig:Model_error}.

\begin{figure*}
    \centering
    \includegraphics[width=0.75\linewidth]{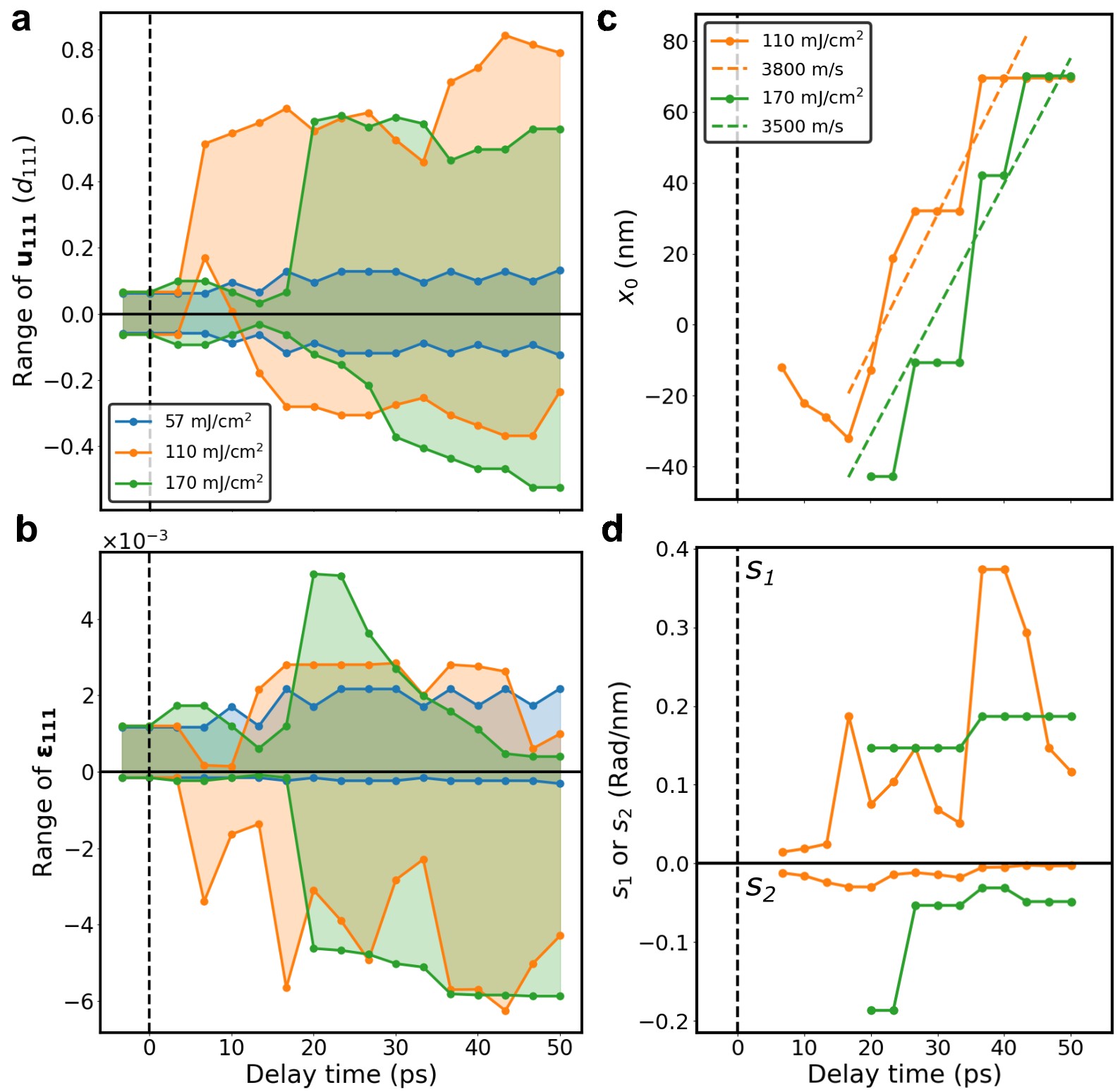}
    \caption{\textbf{The displacement, $\mathbf{u_{111}}$ and strain, $\mathbf{\varepsilon}_{111}$ values at different laser fluences extracted from the 2D model in Fig. \ref{fig:Model}}. The maximum and minimum $\mathbf{u_{111}}$ and $\mathbf{\varepsilon}_{111}$ are displayed in \textbf{a} and \textbf{b}, respectively. The average and standard deviations of $\mathbf{u_{111}}$ and $\mathbf{\varepsilon}_{111}$ are shown in Supplementary Fig. \ref{supp-fig:Strain_disp_values_extra}. The main model variables in Eq. \ref{eq:psi}: \textbf{c} wave peak position, $x_0$; \textbf{d} positive slope, $s_1$, leading to $x_0$, and negative slope, $s_2$, descending from $x_0$. The dashed lines in \textbf{c} show the speed of the boundary propagation.}
    
    \label{fig:Strain_disp_values}
\end{figure*}

In Fig. \ref{fig:Model}a, we observe that the crystal has a homogeneous displacement and relatively low strain at a laser fluence of $\mathrm{57\ mJ/cm^2}$. There is no noticeable splitting in the Bragg peak, which is reflected by the homogeneous displacement in the model. At a laser fluence of $\mathrm{110\ mJ/cm^2}$, shown in Fig. \ref{fig:Model}b, we observe an inhomogeneous phase starting at 6.7 ps, with side-by-side compressed and expanded regions of the crystal. The boundary between the two migrates at the speed of sound (Fig. \ref{fig:Strain_disp_values}c), as the delay time increases so that the compressed region gradually diminishes, leaving an expanded crystal. Interestingly, the strain is mostly negative after 16.7 ps, which could be due to thermal expansion from the region that the heat pulse has already passed through. As the laser fluence is increased to $\mathrm{170\ mJ/cm^2}$, shown in Fig. \ref{fig:Model}c, we also observe the boundary propagating through the crystal at the speed of sound, responsible for the Bragg peak split around 30 ps. This time, neither of the two distinct regions of the crystal with different average lattice parameters is compressed. It may be that the higher heating rate negates the compression, or it is possibly due to the limited spatial resolution \cite{Kim2021a}. The sharper compression wave with a higher compressive strain suggests a more compressed wavefront coinciding with a higher laser fluence, leaving a slightly tensile region after the wave passes through.

Displacement and strain trends can be extracted from the model, shown in Fig. \ref{fig:Strain_disp_values}. The range of $\mathbf{u_{111}}$ is shown in Fig. \ref{fig:Strain_disp_values}a. For all laser fluences, the values increase in magnitude at positive delay times. At $\mathrm{110\ mJ/cm^2}$ and $\mathrm{170\ mJ/cm^2}$ we observe small oscillations in the maximum and minimum $\mathbf{u_{111}}$ with a period of 20 ps. At positive delay times, the average $\mathbf{u_{111}}$ increases with increasing laser fluence (Supplementary Fig. \ref{supp-fig:Strain_disp_values_extra}a). The standard deviation of $\mathbf{u_{111}}$ also increases at positive delay times (Supplementary Fig. \ref{supp-fig:Strain_disp_values_extra}b), demonstrating the increased levels of lattice displacement heterogeneity. 

The range of $\varepsilon_{111}$ is shown in Fig. \ref{fig:Strain_disp_values}b. For $\mathrm{57\ mJ/cm^2}$, we only observe tensile strain, associated with lattice expansion. At $\mathrm{110\ mJ/cm^2}$ and $\mathrm{170\ mJ/cm^2}$, we observe both tensile and compressive strains that persist at positive delay times. At positive delay times, the average $\varepsilon_{111}$ shows an increasing compression, and the standard deviation of $\varepsilon_{111}$ also increases (Supplementary Fig. \ref{supp-fig:Strain_disp_values_extra}c and d, respectively). The latter demonstrates increased levels of strain heterogeneity. 

\section{\label{sec:Discussion}Discussion}

The results can be rationalised with the help of the TTM \cite{Anisimov1974}. The 15 fs laser pulse interacts with the electrons in the Pd nanocrystal, raising them to a high temperature. The electromagnetic skin depth predicts the laser penetration into the metal, determined using Eq. \ref{eq:delta},

\begin{equation}\label{eq:delta}
    \delta = \frac{\lambda_\mathrm{laser}}{4\pi k},
\end{equation}

\noindent where $k_\mathrm{Pd} = 5.09$ is the extinction coefficient for Pd \cite{Johnson1974}, giving $\delta_\mathrm{Pd} = 12.5\ \mathrm{nm}$. Since this is a small fraction of the 180 nm crystal size, the material within this small depth does not contribute very much to the diffracted signal. The electrons within this skin depth will be multi-photon excited to high energy states during the femtosecond pulse, creating a non-equilibrium hot electron distribution. Hot electrons can travel rapidly through the crystal before electron-phonon coupling takes place, creating phonons that heat the lattice on a picosecond timescale. The 120 ps period vibrational mode oscillations in Fig. \ref{fig:Bragg_peak_COM_FWHM_x}a(i) are caused by these phonons travelling back and forth across crystal A. The size of crystal A is around 180 nm, shown in the 2D model (Fig. \ref{fig:Model}). The speed of sound is therefore $180\times10^{-9}\ \mathrm{m}/(120 \times10^{-12}\ \mathrm{s}/2) = 3000\ \mathrm{m\ s^{-1}}$, in agreement with the reported value of $3070\ \mathrm{m\ s^{-1}}$ \cite{Samsonov1968}.

To support these statements, we created a one-dimensional (1D) TTM simulation of a Pd slab using the \texttt{udkm1Dsim} toolbox \cite{Schick2021} (see Supplementary Note \ref{supp-sec:TTMS}). The distribution of electron and lattice temperatures as a function of time and depth into the crystal (Supplementary Fig. \ref{supp-fig:TTM_simulation_maps}a and b) agree well with our interpretation of the experimental data. The 120 ps period strain oscillations are also shown in the 1D TTM simulation (Supplementary Fig. \ref{supp-fig:TTM_simulation_maps}c).

At higher laser fluences, most notably at $\mathrm{110\ mJ/cm^2}$ and $\mathrm{170\ mJ/cm^2}$, we observe Bragg peak splitting most prominently around 30 ps delay time. The peak splitting also reflected the 1D TTM simulation (Supplementary Fig. \ref{supp-fig:TTM_x-ray_simulation}a). Bragg peak splitting has been reported on $\mathrm{SrRuO_3}$ thin films, which is caused by strain variations in different regions following laser excitation \cite{Schick2014}. Similarly, the pronounced splitting we observe corresponds to inhomogeneous strain. This is due to the non-uniform heating of the crystal, which differs from uniform acoustic waves produced by a laser travelling along the depth direction in a thin film \cite{Thomsen1984}. Here, the splitting timescale and partial recovery match heterogeneous electron heating in the TTM: the hot electrons do not fully redistribute within the Pd crystal before transferring their heat to the lattice. This leads to a stress gradient inside the crystal, which drives a sharp strain wave and an inhomogeneous expansion of the lattice (Fig. \ref{fig:Model} and Supplementary Fig. \ref{supp-fig:TTM_simulation_maps}c) as well as diffraction pattern broadening (Fig. \ref{fig:Bragg_peak_COM_FWHM_x}a(ii)-c(ii)) and splitting (Fig. \ref{fig:Bragg_peaks} and Supplementary Fig. \ref{supp-fig:TTM_x-ray_simulation}c).

At the lowest laser fluence used in this study, $\mathrm{57\ mJ/cm^2}$, we did not see signs of splitting in any of the measured Pd crystals, but a clear increase in the FWHM around 30 ps instead. These results generally agree with a previous Au nanocrystal melting study \cite{Clark2015a} that used laser fluences up to $\mathrm{\approx56\ mJ/cm^2}$. For Au, the laser penetration depth is $\delta_\mathrm{Au} = 13.0\ \mathrm{nm}$, assuming $k_\mathrm{Au}=4.91$ \cite{Johnson1974}, similar to the value for Pd. However, there is a large difference in the electron-phonon coupling parameter, with Pd over five times greater than Au \cite{Medvedev2020}, resulting in the electron mean free path of hot electrons being larger in Au than Pd. Therefore, the laser-heated electrons distribute throughout a larger volume of an Au nanocrystal before interacting with the lattice, heating the crystal relatively uniformly and reducing the likelihood of peak broadening or splitting. Additionally, the electronic specific heat capacity, proportional to the electronic density of states at the Fermi level, is about 16 times larger for Pd \cite{Chen1995} than for Au \cite{Lin2008}. For a given energy, the electrons in Au will reach higher temperatures than those in Pd, creating a larger temperature gradient and thus faster thermal transport from the hot electrons to the lattice. These comparisons to Au are further supported by a 1D TTM simulation of Au with comparable strain as the Pd simulation (see Supplementary Note \ref{supp-sec:TTMS}, Supplementary Fig. \ref{supp-fig:TTM_simulation_maps}d-f and Supplementary Fig. \ref{supp-fig:TTM_x-ray_simulation}b).


In our experiment, the expected average temperature rise in the crystal, $\Delta T$, due to one laser pulse can be estimated by $\Delta T = e_{111}/\alpha_\mathrm{Pd}$, where $e_{111}$ is the homogeneous strain and $\alpha_\mathrm{Pd} = 11.8\times10^{-6}\ \mathrm{K^{-1}}$ is the coefficient of linear thermal expansion. The homogeneous strain is defined in Eq. \ref{eq:homo_strain},

\begin{equation}\label{eq:homo_strain}
    e_{111} = \frac{Q_0-Q}{Q_0},
\end{equation}

\noindent where $Q_0 = 2.8038 \textup{~\AA}^{-1}$ is the value at negative delay times, shown in Fig. \ref{fig:Bragg_peak_COM_FWHM_x}a(i). The tensile strain and corresponding maximum rise in temperature are shown in Table \ref{table:strain_temp}. These average temperatures are well below the melting temperature of Pd.

\begin{table}
    \caption{\textbf{Expected homogeneous strain and temperature rise for different laser fluences.}}
    \begin{center}
        \begin{tabular}{|c|c|c|}      
            \hline
            \multicolumn{1}{|c|}{$\mathrm{mJ/cm^2}$} & Homogeneous strain (\%) & Temperature rise (K) \\ \hline
            $57$ & 0.026 & 22 \\
            $110$ & 0.088 & 75 \\
            $170$ & 0.19 & 160 \\
            $230$ & 0.24 & 210 \\ \hline
        \end{tabular}
        \label{table:strain_temp}
    \end{center}
\end{table} 


This compression due to pressure build-up from pulsed laser heating has been previously reported in Pd thin films \cite{Suzana2023b}. For crystal A, in Fig. \ref{fig:Bragg_peak_COM_FWHM_x}, we observe a compression of the average lattice constant between 0 to 20 ps at $\mathrm{170\ mJ/cm^2}$ and $\mathrm{230\ mJ/cm^2}$. Using Eq. \ref{eq:homo_strain}, this corresponds to a compressive strain of $-0.025\ \%$ and $-0.032\ \%$, respectively. This is consistent with a previous study on Pd thin films with a laser fluence of $\mathrm{800\ mJ/cm^2}$, which revealed a maximum compression of $-0.032\ \%$ at 20 ps \cite{Suzana2023b}.

The coherent diffraction results allow the compression to be visualised directly in the model images of Fig. \ref{fig:Model}. The effect is clearest at $\mathrm{110\ mJ/cm^2}$ in Fig. \ref{fig:Model}b and appears to be overwhelmed by larger expansions at higher fluences. It appears that because of the inhomegeneous electron heating, only the front part of the crystal facing the laser expands to create a pressure wave that compresses the back of the crystal, which is also observed by our 1D TTM simulation of Pd in Supplementary Fig. \ref{supp-fig:TTM_simulation_maps}c. As thermal diffusion causes the temperature to equilibrate, the boundary between the expanded and compressed regions is seen to have migrated through the crystal after 50 ps. 

According to Fick's law of diffusion, the heat flow is proportional to the temperature gradient. The solution to the one-dimensional heat equation predicts a Gaussian distribution from an impulse of heat at time $t=0$, with a thermal diffusion width, $w$,

\begin{equation}\label{eq:Fick's}
    w = \sqrt{\frac{2k_Tt}{C_p\rho}},
\end{equation}

\noindent where $k_T$ is the thermal conductivity ($72\ \mathrm{W\ m^{-1}\ K^{-1}}$), $C_p$ is the specific heat capacity ($238\ \mathrm{J\ kg^{-1}\ K^{-1}}$) and $\rho$ is the density ($12\ \mathrm{g\ cm^{-3}}$) \cite{Samsonov1968}. The thermal equilibration time of a 180 nm Pd crystal is therefore expected to be 650 ps, which is longer than the observed elastic wave takes to propagate through the crystal (at the speed of sound), taking 60 ps to travel 180 nm. We conclude that there is a nanoscale stress inhomogeneity after optical excitation in the ultrafast regime, where we have no local thermal equilibrium between electrons and phonons. Hot electron transport alters the energy distribution within the sample, thereby determining the observed strain dynamics. 

\section{\label{sec:Conclusion}Conclusion}
We have presented the results of an optical laser pump and coherent X-ray diffraction probe experiment to reveal nonequilibrium lattice dynamics in individual Pd nanocrystals. These results show laser fluence-dependent changes in the heating of Pd nanocrystals at different delay times. We documented a diffraction peak splitting that occurs around 30 ps delay time, which reveals uneven heating of the nanocrystal. This was not seen in previous studies on individual Au nanocrystals of similar size \cite{Clark2013,Clark2015a}. We attribute the observed inhomogeneity due to hot electrons in Pd having a shorter mean free path and Pd possessing a higher electronic specific heat capacity compared to Au, supported by our 1D TTM simulations. The uneven heating induces a lattice strain distribution, visualised by fitting a coherent diffraction model to the data. Our ability to image the mechanisms involved in heat transfer through lattice strain allows us to gain a better understanding of fundamental heat transport in single nanocrystals. This has applications in predicting reaction rates and preventing thermal degradation in photo-catalysis \cite{Li2020d}.

\section{\label{sec:Methods}Methods}
\subsection{\label{sec:Sample synthesis}Sample synthesis}
Octahedral-shaped palladium nanocrystals (Fig. \ref{fig:SEM}) were synthesised using a seed-mediated approach. The procedure is described here, reproduced from our previous study \cite{Suzana2021}. Nanocrystals were first prepared by combining 0.500 g of cetrimonium bromide (CTAB), 0.5 mL of 0.01 M tetrachloropalladate(II) acid (H$_2$PdCl$_4$), 0.3 mL of 0.01 M sodium iodide (NaI), and 10 mL of nanopure water. This mixture was heated in an oil bath at 95~$^\circ$C for 5 minutes. Subsequently, 200~$\mu$L of 0.04 M ascorbic acid solution was added, and the temperature was maintained at 95~$^\circ$C for 1 hour. For the growth step, 0.360 g of CTAB, 0.250 mL of 0.01 M H$_2$PdCl$_4$, 0.050 mL of 0.001 M NaI, and 9.375 mL of nanopure water were combined and kept at 30~$^\circ$C. After 5 minutes, 80~$\mu$L of the previously prepared Pd seed solution and 250~$\mu$L of 0.04 M ascorbic acid solution were introduced. The reaction was then allowed to proceed at 30~$^\circ$C for 40 hours. Crystalline TiO$_2$ coatings were deposited using a Cambridge Nanotech Savannah 100 system. The deposition was carried out at 500~$^\circ$C under a continuous argon flow of 20 sccm (base pressure approximately 100 mTorr). Titanium isopropoxide (Ti(iPrO)$_4$), heated to 75~$^\circ$C, served as the titanium precursor, while nanopure water at room temperature acted as the oxygen source. The pulse and purge times were 0.1 s and 5 s for Ti(iPrO)$_4$, and 0.01 s and 10 s for H$_2$O. The nanocrystals were drop-cast onto a single-crystal Si wafer and then fixed with tetraethyl orthosilicate (TEOS) \cite{Monteforte2016}. 

\subsection{\label{sec:XFEL experiments}X-ray free-electron laser experiments}
Coherent X-ray diffraction experiments were performed at the Materials Imaging and Dynamics (MID) instrument at the European XFEL \cite{Madsen2021}. All measurements were performed using horizontal scattering geometry. We used a self-seeded X-ray with a bandwidth of $\approx1$ eV centred at 9 keV with 0.67\% transmission. The X-ray was focused to a $4\ \mu\mathrm{m}$ spot size using compound refractive lenses (CRLs). The repetition rate of the X-ray pulses in a train was 2.25 MHz, and the trains were delivered at 10 Hz with 60 pulses per train. At each delay motor position, data for 6 seconds (60 trains) were taken. The exposure time was 20 nanoseconds. Only the first pulse per train was used for the data analysis. The $111$ Bragg peaks of single Pd nanocrystals were measured on an adaptive gain integrating pixel detector (AGIPD) \cite{Allahgholi2019}, with a detector pixel size of $200\ \mu\mathrm{m}$, positioned 4 m from the sample at 2$\theta$ = 35.8 °. The optical laser delivered 15 fs pulses with a central wavelength of 800 nm matching the X-ray pulse pattern. It was focused to a $20\ \mu\mathrm{m}$ spot size at the sample position. All delay measurements, except for those performed on crystal B, were performed stroboscopically, alternating between pumped and unpumped X-ray pulses.

\subsection{\label{sec:Data processing}Data processing}
Calibrated raw data were processed using DAMNIT \cite{DAMNIT}. The intensity was normalised using the X-ray gas monitor (XGM) \cite{Maltezopoulos2019} intensity at each frame before taking the mean of the first pulses of each pumped pulse train at each delay motor position. For each crystal, the detector image was cropped to an array of $128 \times 128$ such that the Bragg peak was in the centre of the array at negative delay times. The cropping region was fixed for each crystal at positive delay times. The mean of a $20 \times 20$ pixel area in the corner of each cropped image was determined as the background level, which was then subtracted from each cropped image.

\subsection{\label{sec:2D Gaussian fitting}Gaussian fitting}
The \texttt{curve\_fit} function from the Python \texttt{scipy.optimize}  sub-package \cite{Virtanen2020} was used to fit the measured Bragg peaks to a 2D Gaussian function. The 2D Gaussian function, which represents the Bragg peak intensity, $I_\mathrm{fit}$, is shown in Eq. \ref{eq:Gaussian},

\begin{equation}\label{eq:Gaussian}
    I_\mathrm{fit}(x,y) = A_0\exp \left( -\frac{1}{2} \left[ \frac{(x - \mu_x)^2}{\sigma_x^2} + \frac{(y - \mu_y)^2}{\sigma_y^2} \right] \right)+O,
\end{equation}

\noindent where $A_0$ is the amplitude, $\mu_{x,y}$ are the pixel coordinates of the Bragg peak centre, $\sigma_{x,y}$ are the widths in pixels of size $p = 200\ \mu$m, and $O$ is the background offset. The Bragg peak position is determined as $Q_{x,y} = \frac{2\pi p}{\lambda d}\mu_{x,y}+\frac{4\pi}{\lambda}\sin{(\theta)}$, where $\lambda = 1.378 \textup{~\AA}$ at 9 keV, $d = 4$ m, $\theta = 17.9\ ^{\circ}$. The FWHM in reciprocal space units is computed as $\mathrm{\Delta Q}_{x,y} = \frac{4\pi p}{\lambda d} \sigma_{x,y}\sqrt{2 \ln 2}$.

The $\chi^2$ error was computed on a $64 \times 64$ array at the centre of the $128 \times 128$ array using Eq. \ref{eq:chi2},

\begin{equation}\label{eq:chi2}
    \chi^2 = \sum_{}^{n} \frac{(\sqrt{I_\mathrm{fit}} - \sqrt{I_\mathrm{measured}})^2}{I_\mathrm{measured}},
\end{equation}

\noindent where $n$ corresponds to each pixel in the $64 \times 64$ array and $I_\mathrm{measured}$ is the measured Bragg peak intensity.

\subsection{\label{sec:Modelling}Modelling of the internal structure of a Pd nanoparticle}
We assume an initial shape of a rounded rectangle to give a Bragg peak similar to the shape shown at negative delay times. This assumption is not unreasonable, as the Pd nanoparticles resemble octahedral shapes, as shown in a previous study \cite{Suzana2021}, thus a rectangle could be representative of its projection in 2D. The shape parameters were optimised with boundary conditions using the \texttt{minimize} function from the Python \texttt{scipy.optimize} sub-package \cite{Virtanen2020}. The correlation between the normalised intensity of the model, $|\mathcal{F}(S(r))|^2$, and the measured normalised intensity was maximised. This was achieved by minimising $1-r$, where $r$ is the Pearson correlation coefficient computed using the \texttt{pearsonr} function from \texttt{scipy.stats} \cite{Virtanen2020}. This metric was used to account for any Bragg peak intensity normalisation uncertainties. $r$ is also shown in Eq. \ref{eq:XC},

\begin{equation} \label{eq:XC}
     r(x,y) = \frac{\sum\limits_{n}(x_n-\bar{x})(y_n-\bar{y})}{\sqrt{\sum\limits_{n}(x_n-\bar{x})^2}\sqrt{\sum\limits_{n}(y_n-\bar{y})^2}},
\end{equation}

\noindent where $x_n$ is the value of a given pixel and delay time, $\bar{x}$ is the mean of the entire array for that delay time, $y_n$ is the value of another pixel and delay time, and $\bar{y}$ is the mean of the entire array at that delay time. The crystal model arrays and their corresponding model diffraction intensities were zero-padded on all sides of the 2D array to $512 \times 512$ pixels. They were then cropped to $128 \times 128$ pixels to match the experimental data for minimising $1-r$. 

The shape of the crystal was considered to be constant throughout the delay time measurements, as the temperature rise in the Pd lattice is far from the melting temperature. The synthesised nanoparticles have very low defect densities; thus, we consider that there is no significant residual strain or lattice displacement in the absence of laser perturbation. The phase $\psi$, shape $s$, and phase ramp $\phi_{\mathrm{ramp}}$ were combined to create a complex crystal function, 

\begin{equation}\label{eq:S}
    S = s(x,y)\,\exp\!\bigl(i\psi(x,x_0,s_1,s_2)\bigr)\,\exp\!\bigl(i\phi_{\mathrm{ramp}}(x,y,\Delta x,\Delta y)\bigr).
\end{equation}

The phase was modelled as a piecewise-linear profile representing a planar acoustic wave propagating through the crystal:

\begin{equation}\label{eq:psi}
    \psi(x,x_0,s_1,s_2) = 
    \begin{cases} 
        s_1(x-x_0), & \quad x \leq x_0, \\[6pt]
        s_2(x-x_0), & \quad x > x_0,        
    \end{cases}
\end{equation}

\noindent where $x$ is the horizontal position in the array, $x_0$ is the peak position of the wave relative to the centre of the crystal, the amplitude is fixed to $\pi$, $s_1$ is the positive slope leading up to $x_0$, and $s_2$ is the negative slope descending from $x_0$. Note that the slopes do not necessarily extend to the edges of the crystal. Parameter values are shown in Fig. \ref{fig:Strain_disp_values}, with additional horizontal and vertical Bragg-peak shifts shown in Supplementary Fig. \ref{supp-fig:Model_parameters}. This phase profile was applied in Fig. \ref{fig:Model}b and c, where peak splitting was observed. 

The experimental intensities were shifted by pixel offsets $\Delta x$ and $\Delta y$ such that their centre of mass was aligned with the centre of the $512 \times 512$ array, using the \texttt{minimize} function from the Python \texttt{scipy.optimize} sub-package \cite{Virtanen2020}. Normalised intensity values below 0.05 were excluded from the centre-of-mass calculation. Following optimisation, this shift was incorporated into the model via a phase ramp,

\begin{equation}\label{eq:phi}
    \phi_{\mathrm{ramp}}(x,y,\Delta x,\Delta y) = 2\pi \left( \frac{\Delta x}{512}\, x \;+\; \frac{\Delta y}{512}\, y \right),
\end{equation}

\noindent such that $|\mathcal{F}\{S\}|^2$ reproduces the noncentred Bragg peak observed in the experiment.

In Fig. \ref{fig:Model}, the displacement along $[111]$ relative to the crystal at negative delay times, $\mathbf{u_{111}}$, can be computed by \cite{Robinson2009},

\begin{equation}\label{eq:disp}
    \mathbf{u_{111}} = \frac{\mathbf{\psi}+\phi_{\mathrm{ramp}}}{|\mathbf{Q}|},
\end{equation}

\noindent where $\mathbf{Q}$ is the scattering vector. The strain relative to the crystal at negative delay times, $\mathbf{\varepsilon_{111}}$, was calculated by \cite{Robinson2009},

\begin{equation}\label{eq:eps_111}
    \mathbf{\varepsilon_{111}} = \nabla(\mathbf{\psi+\phi_{\mathrm{ramp}}})\cdot\frac{\mathbf{Q}}{|\mathbf{Q}|^2},
\end{equation}

\noindent where the phase gradients were determined by taking the derivative of the complex exponential \cite{Yang2022c},

\begin{equation} \label{eq:chain_rule}
    \frac{\partial(\psi+\phi_{\mathrm{ramp}})}{\partial j} = \mathrm{Re}\left(\frac{\partial e^{i(\psi+\phi_{\mathrm{ramp}})}}{\partial j}\bigg/ie^{i(\psi+\phi_{\mathrm{ramp}})}\right),
\end{equation}

\noindent where $j \in \{x, y\}$ in the 2D model. Note that $\psi$ is not differentiable at $x_0$ since it changes from $s_1$ to $s_2$. The optimised shapes and phases were transformed into the laboratory coordinate frame \cite{Yang2019}.

\section{\label{sec:Data availability}Data availability}
The raw data is publicly available at \cite{p2798}.

\section{\label{sec:Code availability}Code availability}
The data analysis was performed on the Maxwell cluster using JupyterHub \cite{Fangohr2020}. The analysis scripts to reproduce the results are publicly available at \cite{zenodo_ref}.

\section{\label{sec:Author contributions}Author contributions}
I.K.R. conceptualized the study; J.W. curated the data; D.Y. performed formal analysis; I.K.R. acquired funding; J.G., L.W., A.F.S., J.D., A.R.-F., J.H., A.Z., U.B., R.S., J.-E.P. A.M., and I.K.R. performed investigation; A.R.-F., J.H., A.Z., U.B., R.S., J.-E.P. and A.M. were associated with methodology; I.K.R. and A.M. was associated with project administration; A.F.S. and I.K.R. acquired resources; J.W., D.Y., and J.G. was associated with the software; I.K.R. supervised the study; J.G., L.W., A.F.S., J.D., A.R.-F., J.H., A.Z., U.B., R.S., J.-E.P. A.M., and I.K.R. performed validation; D.Y. visualised the study; D.Y. wrote the original draft; All authors reviewed and edited the final manuscript.

\section{\label{sec:Competing interests}Competing interests}
The authors declare no competing interests.

\begin{acknowledgments}
We acknowledge European XFEL in Schenefeld, Germany, for the provision of X-ray free-electron laser beamtime at Scientific Instrument MID (Materials Imaging and Dynamics) under proposal number 2798 and would like to thank Mohamed Youssef for his assistance. Samples were provided by Benjamin P. Williams in the lab of Chia-Kuang (Frank) Tsung at Boston College. Work at Brookhaven National Laboratory was supported by the U.S. Department of Energy, Office of Science, Office of Basic Energy Sciences, under Contract No. DESC0012704. Work performed at UCL was supported by EPSRC.
\end{acknowledgments}

\bibliography{library,dataset}

\end{document}